\documentclass{article}
\usepackage{graphicx}

\date{~}

\begin{document}

\title{%
\vskip-18pt~\\
\centerline{Chain Inflation in the Landscape:}
\centerline{``Bubble Bubble Toil and Trouble''}}

\author{Katherine Freese\rlap{}{$^{1}$} and 
Douglas Spolyar\rlap{}{$^{2}$} 
  \\~\\
\small \it ${}^{1}$Michigan Center for Theoretical Physics,
University of Michigan, Ann Arbor, MI 48109
\footnote{ktfreese@umich.edu}\\
\small \it ${}^{2}$Physics Dept., University of California,
Santa Cruz, CA 95060\footnote{dspolyar@physics.ucsc.edu}\\
\vspace{-2\baselineskip} }

\maketitle
 
\begin{abstract} 
  \noindent
  
  In the model of Chain Inflation, a sequential chain of coupled
  scalar fields drives inflation.  We consider a multidimensional
  potential with a large number of bowls, or local minima, separated
  by energy barriers: inflation takes place as the system tunnels from
  the highest energy bowl to another bowl of lower energy, and so on
  until it reaches the zero energy ground state.  Such a scenario can
  be motivated by the many vacua in the stringy landscape, and our
  model can apply to other multidimensional potentials.  The
  ''graceful exit'' problem of Old Inflation is resolved since
  reheating is easily achieved at each stage.  Coupling between the
  fields is crucial to the scenario.  The model is quite generic and
  succeeds for natural couplings and parameters.  Chain inflation
  succeeds for a wide variety of energy scales -- for potentials
  ranging from 10 MeV scale inflation to $10^{16}$ GeV
  scale inflation.

\end{abstract}

\newpage

\section{Introduction}
                
In 1981, Guth \cite{guth} proposed an inflationary
phase of the early universe to solve several glaring paradoxes of the
standard cosmology: the horizon, flatness, and monopole
problems.  During the inflationary epoch, the Friedmann equation
\begin{equation}
H^2 = 8\pi G \rho /3 + k/a^2
\end{equation}
is dominated on the right hand side by a (nearly constant) false
vacuum energy term $\rho \simeq \rho_{vac} \sim$ {\it constant}.  Here
$H$ = $\dot a /a$ is the Hubble parameter and $a$, the scale factor of
the Universe, expands superluminally, $a \sim t^p$ with $p>1$. During
this period, a small causally connected region of the Universe
inflates to a sufficiently large scale (the scale factor must increase
by about $10^{27}$ or roughly 60 $e$-folds of inflation) to
successfully resolve these cosmological shortcomings.  Since the
period of superluminal expansion is an adiabatic process, the
temperature of the universe drops precipitously during this phase.
Hence, it must be followed by a period of thermalization, in which the
vacuum energy density is converted to radiation leading to the
standard cosmology.

Inflationary models can be divided into two categories: Old
Inflation-type models, in which a scalar field tunnels from false to
true vacuum during a first order phase transition, and slowly rolling
inflation models, in which a scalar field rolls down a flat potential
to its minimum.  In either case, the vacuum energy density of the
field before it reaches its minimum drives the inflationary expansion.

In Chain Inflation, we generalize the variety of possible models by
considering a sequential chain of a large number of tunneling and/or
rolling fields, in which reheating relies on the coupling between the
fields.  This chain of multiple tunnelers can be interpreted as a path in a
multidimensional potential landscape, $V(\phi_1, \phi_2, ... ,
\phi_q$), such as may exist in the stringy landscape,  where $q$ is
the number of fields.  One can think of this path as starting at a
bowl (a local minimum) in this multidimensional parameter space, then
moving down to a sequence of bowls of ever lower energy, until one
reaches the ground state with zero potential.  We can model this such
that each drop to a lower bowl is equivalent to one of the fields
tunneling.

{\bf Shortcomings of previous inflationary models:}
Any successful inflationary model must meet two key requirements: 1)
There must be sufficient inflation, and 2) the universe must
thermalize and reheat.  The original Old Inflation model, in which
bubbles of true vacuum nucleate in a false vacuum background,
failed \cite{guthwein} because the interiors of expanding spherical
bubbles of true vacuum cannot thermalize: the ``graceful exit''
problem. Hence this model does not produce a Universe such as our own.
Shortly after the demise of old inflation, models with slowly rolling
fields were proposed \cite{linde,as}.  As a scalar field slowly rolls
down its potential, superluminal expansion is achieved.  Reheating
then takes place successfully as the field oscillates around its
minimum and decays to radiation.  However, slowly rolling models
typically suffer from fine-tuning of their potentials in order that
they be flat enough to provide sufficient inflation and yet not
overproduce density fluctuations; we note that natural inflation
\cite{nat} is a model in which the required small parameters
arise naturally.

{\bf Overcoming the shortcomings with Chain Inflation:} The new
framework of Chain Inflation has several attractive features.  First,
it can resurrect the basic idea of tunneling inflation (as in old
inflation) in that multiple coupled tunneling fields can achieve
graceful exit.  Second, no fine-tuned parameters are required for the
potentials (unlike the case of most slowly-rolling models).  Third,
the model is quite generic; it succeeds for a wide variety of
parameters and couplings. In particular, chain inflation succeeds for
a wide variety of energy scales, ranging from 10 MeV to Grand Unified
scales ($10^{16}$GeV).  Fourth, it relies upon the fact that the
fields are coupled to one another, which, in general, they probably
are.  We will illustrate these features in the paper.  The idea of
taking seriously a model of inflation relying upon hundreds (or more)
of scalar fields was motivated by the large number of vacua in the
string theory landscape.

Unlike old inflation, by having a chain of multiple fields tunneling
sequentially, we are able to fulfill both requirements for inflation:
sixty e-folds of inflation as well as reheating.  The key element is
that no single stage of inflation is responsible for much inflation.
Each stage of inflation gives rise to only a fraction of an e-fold, and
it is only due to the large number of stages of tunneling that the
universe inflates sufficiently.  Graceful exit from inflation is
obtained by coupling the fields together.  Once a field has tunneled
to its true vacuum, its coupling to secondary field(s) causes a change
in the nucleation rate in the secondary field(s); the rapid tunneling
of the secondary field(s) to the true vacuum allows bubble percolation
and reheating.  A chain of such tunneling events, each catalyzed by a
previous tunneler, leads to a homogeneous hot universe.

In chain inflation, the possible range of the energy scales of the
potentials can vary between $10^{16}$ GeV down to any energy scale
that allows sufficient reheating and baryogenesis.  Thus
the potential can have an energy scale possibly as low as 10 MeV, so
that the universe can reheat to that energy scale and still experience
ordinary Big Bang Nucleosynthesis\footnote{The lower bound on the
  energy scale of inflation would be set by baryogenesis, which is
  currently not understood. For example, if baryogenesis is found to
  take place at the electroweak scale, then the lower bound on the
  chain inflation potential would be roughly TeV scale.  However, as
  we don't currently know how baryogenesis works, we can contemplate
  even lower energy scales.}.  In short, whereas most traditional
rolling models of inflation require large energy scales, often above a
Planck scale, in Chain Inflation it is possible for all the potentials
to have much lower energies, e.g., they can all have TeV scales.

In order to illustrate the basic scenario, we first consider a chain
of $q$ tunneling fields, all with the same parameters.  In this simple
picture, once the first field tunnels, it catalyzes the second field
to tunnel (where the second field without the coupling to the first
would remain in the metastable vacuum), the second field catalyzes the
third field to tunnel, etc.  The key point is that the universe
inflates a fraction of an e-fold at each stage, and the coupling
ensures that the sequence of tunneling events takes place.  After
discussing the simplest variant of the chain in which each field is
coupled only to two others (the previous and subsequent ones in the
chain), we turn to a more generic situation in which the potentials
are allowed to have a variety of parameters and couplings exist
between multiple fields.  The same basic behavior -- a sequence of
tunneling events -- ensues, as the universe chooses a path to the
ground state.  We also will assume that the ground state is
Minkowski space ($V=0$).

Whether or not reheating is successful for a tunneling field depends
crucially on the ratio between the expansion rate $H$ and the
bubble-nucleation rate $\Gamma$ per unit volume.  We can define the
ratio
\begin{equation}
\label{eq:beta}
\beta = \Gamma/H^4 .
\end{equation}
Roughly speaking, if $\beta <<1$, bubble nucleation is rare, and the
universe stays trapped in the false vacuum and inflates.  If $\beta
\geq O(1)$, on the other hand, bubble nucleation is rapid, there is no
further inflation, and the phase transition successfully completes via
bubble percolation and thermalization.  Old inflation, which had a
single tunneling scalar field, required $\beta << 1$, so that 60
efolds could be achieved; then for a $\beta$ that is constant in time,
it became impossible for nucleation ever to successfully complete.
Double field inflation, proposed in 1990 by Adams and Freese
\cite{adamsfreese} solved this problem with a time dependent
nucleation rate for a single tunneling field, so that $\beta$ started
out small and the universe inflated; later due to coupling to a
rolling field, $\beta$ for the tunneling field became large so that
the phase transition suddenly took place and completed throughout the
universe.  In this paper we use some of the features of this Double
Field model. We couple a series of tunneling fields together in such a
way that each member of the chain catalyzes the next to rapidly
nucleate and thermalize.

As a toy model, we will consider many scalar fields, each of which has
an asymmetric double well potential.  We do not mean to imply that
this is a perfect model for the real potentials of these fields, but
merely an illustrative example.  Of course the shape of the
multi-dimensional potential is far more complicated than our simple
picture.  In addition, we do not deal with anti-de Sitter vacua.
Despite these reservations, our model has the advantage that it is
quite generic.  The basic features of the model should survive if one
considers more complicated potentials.  The model succeeds for a wide
variety of parameters and couplings.  We allow arbitrary couplings
between the fields and find that their potentials do not need to be
fine-tuned.  Hence the model has the positive feature that it succeeds
for potentials that are natural in the context of field theory.  To be
succinct, if we take a large number of scalar fields which are coupled
together and have natural potentials (two criteria which are
reasonable in the context of field theory and possibly also in the
context of a landscape), we have the necessary conditions for
inflation.

The generic feature of our model is this idea that, in a complicated
landscape, the system drops through a series of bowls by tunneling
from one to another, until it hits the ground state.  We will consider
a specific form of couplings between fields in order to allow us to
perform calculations and get reasonable estimates.  However, we
emphasize that the general idea of tunneling from bowl to bowl does
{\it not} rely on the choices of potential and couplings that we use
in order to present a concrete toy model.  Our model relies only on
the one general feature of tunneling from bowl to bowl, regardless of
the detailed form of the potential.

Since our treatment is entirely field theoretical, this work does not
rely on the existence of any stringy landscape.  This model may be
relevant to many systems with multidimensional potentials, such as
condensed matter systems or in particle physics.  Previous authors
\cite{liddle}, \cite{kaloper} and \cite{easther} have considered
multiple rolling fields, without any tunneling fields present; our
work differs in that we generically have tunneling fields in the model
and in that the couplings between the fields are key to the success of
the model.  We note that in the future it would also be interesting to
combine a number of rolling as well as tunneling fields. We also plan
to shortly publish a paper considering inflation from a chain of vacua
in a single field: one might imagine a tilted cosine, and in fact we
find that chain inflation can succeed with the QCD axion \cite{fls}.
The case of a large number of uncoupled scalar fields was previously
considered by \cite{liddle,kaloper}.  A chain of rolling fields as the
source of inflation was considered by Easther \cite{easther} in a
model he dubbed ``folded inflation.''  Our work, which focuses on
tunneling fields, is hence complementary to his paper.

One important issue we have not yet dealt with in this paper is the
question of density fluctuations.  One of the outstanding successes of
rolling models in inflation is that they are able to produce density
fluctuations with a scale invariant spectrum.  However, unless their
potentials are fine-tuned, they tend to overproduce the amplitude of
the density fluctuations (except in a few models such as natural
inflation \cite{nat}).  In order for Chain Inflation to be a viable
competitor to rolling models, it too should produce density
fluctuations of the right spectrum and amplitude.  Towards the end of
the paper (in the discussion section) we present a partial discussion
of the expected results, but far more work must be done on this
important issue\footnote{In the worst case, one might argue that, as
  long as Chain Inflation does not overproduce the perturbations, they
  might be produced elsewhere in the universe, but it would be
  preferable to generate them during inflation.}.

We begin with a review of tunneling in double-well potentials and the
failures of old inflation in Section II. In Section III, we review the
Double Field model, which revives the idea of old inflation.  In
Section IV, we present the idea of Chain Inflation: we start with the
toy model of identical tunnelers in which all the fields have
potentials with identical parameters, and then generalize to coupled
tunnelers with different parameters.  In Section V, we present several
variants on the idea of a chain of tunnelers. In Section VI there is a
discussion (including on the issue of density perturbations) and we
conclude in Section VII.

\section{Review of Tunneling in Double Well Potential and Old Inflation} 
We consider a quantum field theory of a scalar field with a Lagrangian
of the form
\begin{equation}
\label{eq:lag}
{\cal L} = {1 \over 2}(\partial_{\mu}\phi)(\partial^{\mu} \phi) 
- V(\phi) , 
\end{equation} 
where $V (\phi)$ is an asymmetric potential with metastable minimum
$\phi_-$ and absolute minimum $\phi_+$ (see Fig. 1).  The energy
difference between the vacua is $\epsilon$.  Bubbles of true vacuum
($\phi_+$) expand into a false vacuum ($\phi_-$) background.

\begin{figure}
\centerline{\includegraphics[width=5.5in]{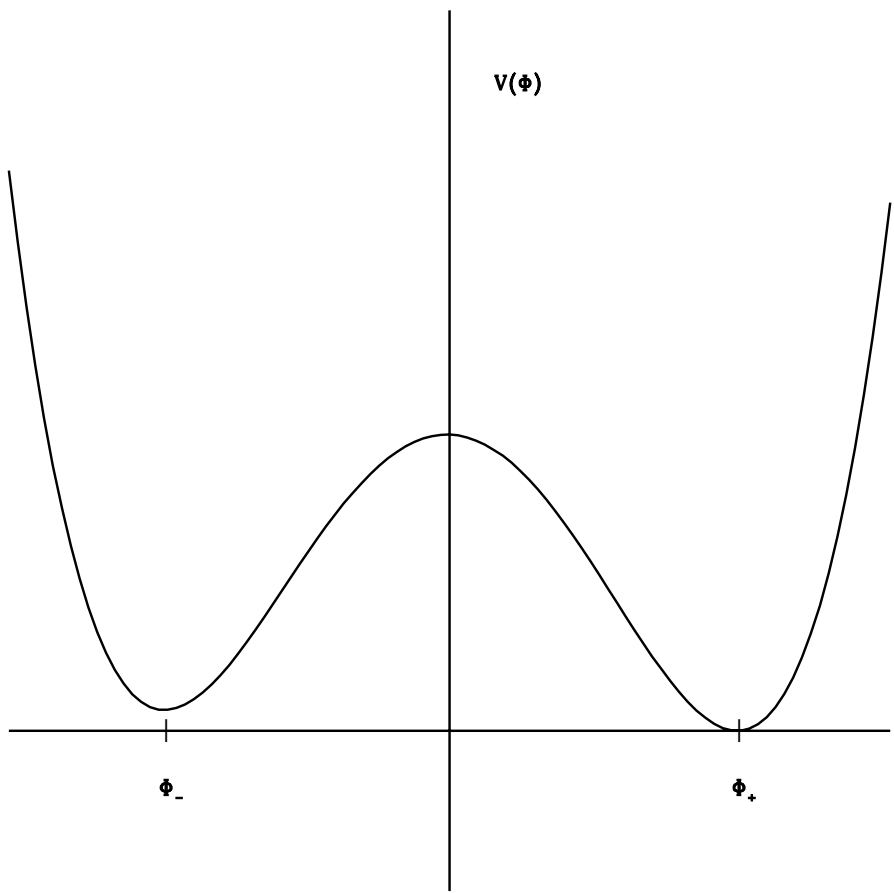}}
\caption{Potential energy density of tunneling field $\phi$ as 
  a function of field strength. The energy difference $\epsilon$
  between the false vacuum (at $\phi_1=-a$) and the true vacuum (at
  $\phi_+ = +a$) provides the vacuum energy density for inflation.}
\end{figure}

In the zero-temperature limit, the nucleation rate $\Gamma$ per unit
volume for producing bubbles of true vacuum in the sea of false vacuum
through quantum tunneling has the form \cite{callan,coleman}
\begin{equation}
\label{eq:tunrate}
\Gamma(t) = A e^{-S_E}
\end{equation}
where $S_E$ is the Euclidean action and where $A$ is a determinantal
factor which is generally the energy scale $\epsilon$ of the phase
transition\footnote{We
  note that we do not need to include gravitational effects
  \cite{deLuccia} as they would only be relevant for bubbles
  comparable to the horizon size, whereas the bubbles produced here
  are much smaller.}. Guth and Weinberg have shown that the
probability of a point remaining in a false deSitter vacuum is
approximately
\begin{equation}
\label{eq:probds}
p(t) \sim {\rm exp}({-{4 \pi \over 3}\beta H t})
\end{equation}
where the dimensionless quantity $\beta$ is defined by
\begin{equation}
\beta \equiv {\Gamma \over H^4 }.
\end{equation}
Writing Eq.(\ref{eq:probds}) as $p(t) \sim {\rm exp}(-t/\tau)$, we can
estimate the lifetime of the field in the metastable vacuum as
roughly\footnote{There will be a distribution around this typical
  value, as we will discuss further below.}
\begin{equation}
\label{eq:lifetime}
\tau = {3 \over 4 \pi H \beta} = {3 \over 4 \pi} {H^3 \over \Gamma}.
\end{equation}

For definiteness, we consider asymmetric
double well potentials as in Figure (1),
\begin{equation}
\label{eq:doublewell}
V(\phi) = {1 \over 4} \lambda (\phi^2 - a^2)^2 - {\epsilon
\over 2a}
(\phi -a ) .
\end{equation}
To leading order, the metastable minimum is at $\phi = -a$ and the
absolute minimum at $\phi = +a$.  The energy difference between minima
is $\epsilon$.  Throughout the paper we will assume that $\lambda$
is not too different from 1, as it is the most natural value of this
parameter.

In the thin wall limit\footnote{The thin-wall approximation breaks
  down for much of the parameter space we are considering. However, as
  illustrated in the Discussion Section below, the results of the
  paper would be relatively unchanged if we were to perform a more
  careful study in which this approximation is not made.  Hence we
  proceed cautiously using the thin wall limit.}, the Euclidean action is
\begin{equation}
\label{eq:eucact}
S_E = 64{\pi^2 \over 6}  {\lambda^2 a^{12} \over \epsilon^3 }.
\end{equation}

When $\beta << 1$ (low nucleation efficiency), the phase transition
proceeds slowly. The field remains in its metastable minimum and the
universe inflates.  However, as long as $\beta$ is small, the phase
transition cannot complete.  Bubbles of true vacuum do periodically
nucleate in various places in the universe, but their production is rare
and sporadic.  The rate of filling
the universe with true vacuum cannot keep up with the exponential
expansion of the false vacuum.  Percolation of true vacuum bubbles
does not take place.  The interiors of individual bubbles or the small
groups of bubbles that are able to form are unable to thermalize.  The
latent heat of the phase transition resides entirely in the
kinetic energy of the bubble walls and cannot thermalize the interior.
The universe ends up looking like ``Swiss Cheese'', with isolated
empty bubbles of true vacuum of various sizes unable to
find one another in the background of false vacuum.  

Old inflation, which has constant (time-independent) small $\beta$, fails.
In old inflation, the universe can easily grow the requisite 60
e-foldings, but reheating never takes place.

In the opposite limit of $\beta > O(1)$ (high nucleation efficiency),
the phase transition proceeds very rapidly. If there were only one
scalar field, it would not remain in the false vacuum long enough to
give rise to sufficient inflation, though the phase transition would
complete, leading to percolation and thermalization.  Guth and
Weinberg \cite{guthwein} as well as Turner, Weinberg, and Widrow
\cite{tww} calculated that a critical value of
\begin{equation}
\label{eq:betacrit}
\beta \geq \beta_{crit} = 9/4\pi
\end{equation}
is required in order for percolation and thermalization to be
achieved.  In this paper, we consider a chain of multiple tunneling
fields rather than a single field. Each of these fields will develop a
value of $\beta$ in excess of the critical value required for
percolation.

The amount that the universe inflates before the phase transition
completes varies inversely to the parameter $\beta$.  The
number of e-foldings due to tunneling of a single scalar field is
\begin{equation}
\label{eq:ni}
N = \int H dt \sim H \tau . 
\end{equation}
Using the first equality in Eq.(\ref{eq:lifetime}) for the lifetime
$\tau$, we then have
\begin{equation}
N = {3 \over 4 \pi \beta} .
\end{equation}
Sufficient inflation requires the total number of e-foldings to
satisfy 
\begin{equation}
\label{eq:totreq}
N_{tot}>60.
\end{equation}
Old inflation can easily achieve this requirement with a single scalar
field for a wide variety of parameters. For a double-well potential in
the thin wall limit, the only relevant quantity in the potential for
determining the total amount of inflation is the ratio $\epsilon /
a^4$.  Hence old inflation can take place at a variety of energy
scales, and can be equally as successful at a TeV as at $10^{16}$GeV.

However, using Eq.(\ref{eq:betacrit}), we see that percolation and
thermalization of any single field require
\begin{equation}
\label{eq:ncrit}
N \leq N_{crit} = 1/3.
\end{equation} 
For the case of inflation with a single scalar field, $N_{tot}=N$ and
obviously one cannot simultaneously satisfy both criteria in
Eqs.(\ref{eq:totreq}) and (\ref{eq:ncrit}).  Hence with
time-independent $N$ and $\beta$, old inflation fails.  However, as we
will illustrate in the next section, both of these criteria can be
satisfied with a time-dependent nucleation rate, in which case $\beta$
and $N$ change with time.

This transition from small $\beta$ (large N) to large $\beta$ (small
N) can be achieved by a small change in the parameters in the
potential.  Because of the exponential dependence in
Eq.(\ref{eq:eucact}) on the Euclidean action, the values of $\beta$
and $N$ are extremely sensitive to the parameters in the potential, in
this case to the ratio $a/\epsilon^{1/4}$.  Because the tunneling rate
is so sensitive to this ratio, the transition from a field that would
remain for an inordinate amount of time in the false vacuum to a field
that tunnels rapidly requires only a small change in the ratio of
$a/\epsilon^{1/4}$.  
As discussed later, this ratio only needs to change
by a few percent  to cause a transition from slow to rapid tunneling 
for a wide range  of energy scales from GUT to TeV.

It is interesting that potentials which have comparable values of
$\epsilon$ and $a^4$ are near the borderline between rapid transitions
and no transition.  We will take advantage of this fact in this paper.
Such potentials with parameter $\epsilon \sim O(a^4)$ are quite
natural.

\section{Double Field Inflation: Time-Dependent Nucleation Rate}

Adams and Freese \cite{adamsfreese} solved the problem of old
inflation for a single tunneling field by using a time-dependent
nucleation rate and hence a time-dependent $\beta$ (see also \cite{Linde:1990gz}).  The ideal
time-dependence of $\beta$ would be a step-function. In that case,
$\beta$ is  initially small so that one obtains the required 60 e-foldings of
inflation, followed by a sudden transition to a large value of $\beta$
so that all of the universe goes from false to true vacuum at once. Then
all the bubbles of true vacuum are roughly of the same size,
and they are able to percolate and thermalize (leaving no Swiss
Cheese).  Adams and Freese obtained the appropriate time dependence of
$\beta$ by coupling a tunneling inflaton field $\phi$ to a rolling
field $\psi$.

They took the total potential for the two fields to be
\begin{equation}
V_{tot} = V_1(\phi) + V_2(\psi) + V_{int}(\phi,\psi) ,
\end{equation}
where the inflaton field $\phi$ is a tunneling field with potential
$V_1(\phi)$ given by the form of Eq.(\ref{eq:doublewell}), and $\psi$
is a rolling field with potential $V_2(\psi)$.  The purpose of the
rolling field is to catalyze a change in the tunneling rate of the
tunneling inflaton.  As interaction, they considered
\begin{equation}
\label{eq:vint}
V_{int}(\phi,\psi) = - \gamma (\phi -a) \psi^3 .
\end{equation}
Clearly many other forms of the potential are possible (e.g., $V_{int}
\sim \phi^2 \psi^2$), but the resulting behavior should be
qualitatively the same.  Due to this interaction, (in the thin wall
limit) bubbles will nucleate at a rate given by Eq.(\ref{eq:tunrate}),
where the effective energy difference between the vacua is given by
\begin{equation}
\label{eq:epsiloneff}
\epsilon_{eff} = \epsilon + 2 a \gamma \psi^3.
\end{equation}
The tunneling rate is practically zero when the rolling
field $\psi$ is at the top of its potential 
and abruptly becomes very large as $\psi$ approaches the 
minimum of its potential.

Initially, when the value of the rolling field $\psi \sim 0$ is small, 
the interaction term $V_{int}(\phi,0) \sim 0$ is negligible, and the
energy difference $\epsilon_{eff} \sim \epsilon$ 
between the two minima is small
enough that the field is stable in the false vacuum.  The parameter 
$\beta$ is small
and the universe inflates. Then, as $\psi$ rolls to the bottom of its
potential, the energy difference $\epsilon_{eff}$ between the minima
of the tunneling field grows and the tunneling rate given by
Eqs.(\ref{eq:tunrate}) and (\ref{eq:eucact}) climbs sharply due
to the exponential dependence on the value of $\epsilon_{eff}^3$.
Suddenly $\beta$ becomes larger than the critical value for nucleation
and true vacuum bubbles of the tunneling field appear everywhere at
once.  The universe successfully completes the phase transition.

The strength of this Double Field model is that it saves old
inflation.  The bubble nucleation is sudden and occurs throughout the
universe, so that it can percolate and thermalize.  
In the Double Field model, the rolling field must be flat
(just as in new inflation)
in order for the tunneling field to remain in its false vacuum
for a sufficiently long time and in order
not to overproduce density fluctuations.  
To explain this flatness, Freese \cite{pngb} used a
pseudo-Nambu Goldstone boson (shift symmetry), similar to the model of
natural inflation \cite{nat,natural2}, but then found that the
parameters of the tunneling inflaton potential needed to be fine-tuned
as well.

An alternate approach to obtaining a time-dependent $\beta$ was
pursued in the models of Extended \cite{lastein} and Hyperextended
Inflation \cite{lsb,steinacc}, in which the Hubble constant becomes
time-dependent due to Brans-Dicke gravity \cite{bransdicke}.
However, it has been shown \cite{lidwands,tww} that most versions
of these models are ruled out due to overproduction
of big bubbles in conflict with microwave background \cite{WMAP}
and oher data.

In this paper we preserve some features of the original Double
Field Inflation model.  We couple fields together so as to
modify the value of $\epsilon_{eff}$ in the double well potential and
thereby modify the nucleation rate. Here, however, no individual
tunneling stage is responsible for more than a fraction of an
e-folding, and no fine-tuning of potentials is required. In the
simplest variant of our model, only tunneling fields are implemented,
though some of these may be replaced by rolling fields instead.

\section{Chain Inflation}

The model of Chain Inflation relies on a chain of tunneling fields.
All the fields start out in their false vacua.  The chain is set off
by a single field tunneling to its true vacuum; this tunneling event
then catalyzes a chain of tunneling events of the other fields.  In
the context of a landscape, where there is a multidimensional
potential for a large number of scalar fields, one can think of Chain
Inflation in the following way.  At some place in the universe, the
fields start off in a bowl (metastable minimum) of some energy.  Then,
the tunneling of one field in our chain model is equivalent to moving
to a bowl of slightly lower energy.  In the chain picture, each
tunneling event provides a small amount of inflation (less than one
e-fold) but, in the end, by the time the fields reach their collective
ground state (which we take to be $V=0$), more than 60 e-folds have
taken place.

We will begin by discussing an extremely simplified version of the
chain.  We will have the fields tunnel to their true vacua one at a
time: first one field tunnels (i.e. bubbles of its true vacuum
nucleate), then it catalyzes the second field to nucleate, which
catalzyes the third field to nucleate etc.  Each field couples only to
two others: the prior and subsequent ones in the chain.  This
over-simplified model is like a set of dominoes: once the first domino
falls, all the rest subsequently follow.  After we discuss this
extremely simplified model, we will generalize to additional couplings
between the fields and a variety of parameters for the potentials and
couplings.

Our model relies only on one general feature, of tunneling from bowl to
bowl, regardless of the detailed form of the potential.  Chain
Inflation will happen for a system of scalar fields in a
multidimensional potential, which tunnel from bowl to bowl.  Our
particular choice of couplings allow simple calculations so as to
obtain reasonable estimates. However, the coupling certainly need not
be of the form presented here.  Indeed, all the fields could couple to
each in other in a complicated fashion as long as the couplings
preserve the bowl-type of structure in a multidimensional landscape.

As our toy model, we take all the fields to have double-well
potentials.  The basic features of the model generalize to other
choices of potential.  As a concrete example, we will start with a
particular form of the interaction which is linear in one of the
fields and cubic in the other (simply because it makes the algebra
easy).  Qualitatively, the particular choice of interaction term is
not important, and other choices such as interactions which are
quadratic in both fields, would produce the same behavior of the chain
model.  Other choices of potential or changes in the potential could
equally well drive a time-dependent nucleation rate that allows
inflation as well as percolation.  In this paper, we use a
time-dependence of the (effective) energy difference between minima to
drive a time-dependent nucleation rate.  Alternatively, one could use
a time-changing barrier height or time-dependent potential width to
achieve the same time-dependence of the nucleation rate and hence of
$\beta$. There are many ways to achieve the same effect, and we have
chosen a cubic coupling influencing the value of the energy difference
as a concrete example.

The total potential for the system is
\begin{equation}
\label{eq:vtot}
V_{tot}(\phi_{1},\phi_{2}, ... ,\phi_{q}) 
= \sum_i V_{tot,i} = \sum_i[V_i(\phi_i) + V_{i,i-1}]
\end{equation}
where $0 < i \leq q$.  We take asymmetric double-well potentials
\begin{equation}
V_i(\phi_i) = {1\over 4}
\lambda_i(\phi_i^2 - a_i^2)^2 - {\epsilon_i \over 2 a_i}
(\phi_i - a_i)
\end{equation}
and, as a simple example, for interactions between the fields we take
\begin{equation}
\label{eq:vinti}
V_{i,i-1} = - {\gamma_{i,i-1} \over 16}
(\phi_{i} - a_{i}) (\phi_{i-1} + a_{i-1})^3 .
\end{equation}
The first field in the chain must be treated individually and is
discussed in the next section below.  All the fields start out in
their false vacua, $\phi_{i, initial} = -a_i$.  One after the next, in
a chain, they tunnel to their true vacua at $\phi_{i,final} = + a_i$.
After $i\!\!-\!\!2$ of them have tunneled to their true vacua, the effective
energy difference between minima for field $i$ is given by
\begin{eqnarray}
\label{eq:effdiff}
\epsilon_{eff} \!\! & =  V_{tot}[\phi_{0}=a_{0},...,
\phi_{i-2} = a_{i-2}, \phi_{i-1}=
\mp a_{i-1}, \phi_{i}=-a_{i},\phi_{i+1}=-a_{i+1}, ... , \phi_q = - a_{q}]
\nonumber\\
&\,\,- V_{tot}[\phi_{0}=a_{0},...,\phi_{i-2} = a_{i-2},\phi_{i-1}=
\mp a_{i-1}, \phi_{i}=+a_{i},\phi_{i+1}=-a_{i+1}, ... , \phi_q = - a_{q}]
\nonumber\\
&= \epsilon_i + {1 \over 8} a_i \gamma_{i,i-1} (\phi_{i-1}
+ a_{i-1})^3 - \gamma_{i,i+1} a_i^3 a_{i+1}. 
\label{eq:last}
\end{eqnarray}
The potential in Eq.(\ref{eq:effdiff}) is evaluated at $\phi_{i-1} = -
a_{i-1}$ when the $(i\!\!-\!\!1)$-th field is in its false vacuum, and at
$\phi_{i-1} = + a_{i-1}$ when the $(i\!\!-\!\!1)$-th field is in its
true vacuum.

The last term in Eq.(\ref{eq:last}) arises due to the fact that, once
the field $i$ tunnels to its true minimum, its interaction with field
$i\!\!+\!\!1$ is nonzero, i.e., the field $i$ tunnels to the location
in the landscape where its interaction with the next field in the
chain is active.  Some of the energy $\epsilon_i$ must go into this
interaction energy rather than into the energy of the true vacuum
bubbles. Hence, the interaction with the next field in the chain
enters with a minus sign in Eq.  (\ref{eq:effdiff}).  This {\it
  decreases} the energy difference between the vacua, and makes
tunneling of field $i$ to its true vacuum more difficult (lowers the
tunneling rate).

However, it is still true that each field in the chain successfully
catalyzes the next one to nucleate.  When field $i\!\!-\!\!1$ tunnels to its
true minimum, it increases the tunneling rate of the $i$-th field to
the point where it too tunnels to its minimum.  At first, when
$\phi_{i-1, initial} = -a_{i-1}$ is in its false vacuum, there is no
interaction with $\phi_i$ and
\begin{equation}
\label{eq:epsinit}
\epsilon_{eff,i, initial} = \epsilon_i  - \gamma_{i,i+1} a_i^3 a_{i+1}.
\end{equation}
Then, once the $(i\!\!-\!\!1)$-th field tunnels to its true vacuum at
$\phi_{i-1, final} = + a_{i-1}$, the interaction with field $\phi_i$ turns
on and
\begin{equation}
\label{eq:epsfin}
\epsilon_{eff,i,final} = \epsilon_i +  \gamma_{i,i-1} a_{i-1}^3 a_i  
- \gamma_{i,i+1} a_i^3 a_{i+1}.
\end{equation}
This positive change in the energy difference by an amount
$\gamma_{i,i-1} a_{i-1}^3 a_i$ (the interaction energy between
fields $i\!\!-\!\!1$ and $i$)  is enough to increase the
tunneling rate for field $i$ to the point where bubbles of its true
vacuum nucleate throughout, allowing percolation and thermalization of
these bubbles of field $i$.  

In the language of a landscape, due to the tunneling of field $i\!\!-\!\!1$,
the system moves to a different location in the multi-dimensional
potential.  It starts in a bowl with negligible tunneling rate
for field $i$ from false to true vacuum, and moves
to another bowl from which the field
$i$ can easily tunnel to its minimum.

Let us consider two of the fields in the chain to illustrate how this
works.  We will assume that the first field in the chain (subscript
$0$) has already tunneled to its minimum.  Then the potential terms
felt by fields $1$ and $2$ are as follows:
\begin{eqnarray}
&V_{0,1}  =  -{1 \over 2} \gamma_{0,1}a_0^{3} (\phi_1 - a_1) ,\\
&V_{1} = {\lambda_1\over 4} (\phi_1^{2} - a_{1}^{2})^{2} 
- {\epsilon_{1} \over 2 a_{1}} (\phi_1 - a_{1}) ,\\
&V_{1,2} = - {1 \over 16}\gamma_{1,2} (\phi_{2} - a_{2})(\phi_{1} + a_{1})^{3} ,\\
&V_{2} = {\lambda_{2}\over 4} (\phi_2^{2} - a_{2}^{2})^{2} 
- {\epsilon_{2} \over 2 a_{2}} (\phi_2 - a_{2}) ,\\
&V_{2,3} = - {1 \over 16} \gamma_{2,3} (\phi_{3}-a_{3})(\phi_2+a_{2})^{3} .
\label{eq:terms}
\end{eqnarray}
These are all the terms involving fields 1 and 2.  In the last term,
we will set $\phi_{3}= -a_{3},$ as we are interested in the case (the
location in the landscape) where the third field is still in its false
minimum.  In addition, we can ignore all terms proportional to
$\lambda_i$, since these terms are zero when evaluated at $\phi_i = \pm
a_i$.  Hence, if we include all the terms involving fields 1 and 2 and
ignoring terms with $\lambda_i$, we have
\begin{equation}
\label{eq:veff}
V_{eff}(\phi_1,\phi_2) 
= {-1 \over 2} \bigl( {\epsilon_1 \over a_1} + \gamma_{0,1} a_0^3
\bigr) (\phi_1 - a_1) - {1 \over 2} {\epsilon_2 \over a_2}(\phi_2 - a_2)
- {\gamma_{1,2} \over 16} (\phi_2 - a_2) (\phi_1 + a_1)^3 
+ {\gamma_{2,3} \over 8} a_3 (\phi_2 + a_2)^3 .
\end{equation}

Initially, we start out with $\phi_{1, initial} = -a_1$ and $\phi_{2,
  initial} = -a_2,$.  At these initial values, there is no interaction
between fields 1 and 2, i.e., $V_{1,2}(-a_1,-a_2) = 0$.  Then the first
field tunnels, so that $\phi_1 \rightarrow a_1$.  Now the interaction
term between fields 1 and 2 becomes nonzero so that $\epsilon_{2,eff}$
increases by an amount $\gamma_{2,1} a_2 a_1^3$, following
Eq.(\ref{eq:epsfin}).  Now the tunneling rate in Eq.(\ref{eq:tunrate})
is increased to the point where the second field tunnels and reheats.
We require that $\beta_{i,final} > \beta_{crit} = 9/4\pi$.

One can describe this behavior in the language of a multidimensional
landscape as follows.  There are four bowls, or local minima of the
multidimensional potential: (1) at point A, $\phi_1 = - a_1, \phi_2 =
-a_2$, (2) at point B, $\phi_1 = +a_1$ and $\phi_2 = -a_2$, (3) at
point C, $\phi_1 = -a_1$ and $\phi_2 = +a_2$, and (4) at point D,
$\phi_1 = +a_1, \phi_2 = +a_2$.  Table I illustrates the values for
$V_{eff}(\phi_1=\pm a,\phi_2 = \pm a)$ in these four bowls.
We will show that the path taken in this multidimensional parameter
space is $A \rightarrow B \rightarrow D$.

\begin{table}[tbp]
\begin{tabular}{|l|c|c|c|c|c|l|}
\hline
Bowl & $\phi_1$ & $\phi_2$ & $V_{eff}$ \\
\cline{2-4}
\hline
A & $-a_1$ & $-a_2$ & $\epsilon_1 + \gamma_{0,1} a_0^3 a_1 + \epsilon_2$ \\
\hline
B & $a_1$ & $- a_2$ & $\epsilon_2 + \gamma_{1,2} a_1^3 a_2$ \\
\hline
C & $-a_1$ & $a_2$ & $\epsilon_1 +\gamma_{0,1} a_0^3 a_1+\gamma_{2,3} a_2^3 a_3$ \\
\hline
D & $a_1$ & $a_2$ & $\gamma_{2,3} a_2^3 a_3$ \\
\hline
\end{tabular}
\caption{All potential terms for fields 1 and 2 in the chain,
assuming that field 0 is in its true vacuum and field 3 is still
in its metastable false vacuum.
}
\end{table}

The highest energy bowl is at point A, where both fields are in their
false vacua; this is the starting point of the system.  First, field
$\phi_1$ tunnels from $-a_1$ to $+a_1$ and the system moves from bowl
A to bowl B.  The energy difference between bowls A and B is typically
higher than the energy difference between bowls A and C. Thus, $A
\rightarrow B$ has a higher tunneling rate and provides the preferred
path\footnote{We will presently comment on parameter choices for which
  this is not the preferred path.}.  After the system has taken the
path from bowl A to bowl B in the landscape, the second field $\phi_2$
can now easily tunnel from $-a_2$ to $+a_2$.  The system reaches bowl
D, the bowl with the lowest energy. [As we do not want to end up with
  a cosmological constant, we will in fact subtract off the energy of
  the lowest bowl everywhere.]

In this language of the bowls, the statement that the tunneling of the
first field catalyzes the tunneling of the second field can be given
the following interpretation.  We can see that the first field
catalyzes the second one to tunnel.  The tunneling rate of the second
field via $A \rightarrow C$ (in the location of the landscape
where the first field has not yet tunneled to its true vacuum) is
slow and typically does not take place, with
\begin{equation}
\epsilon_{2,eff,initial} = E_A - E_C = \epsilon_2 - \gamma_{2,3} a_2^3 a_3  
\,\,;
\end{equation}
the system does not choose this path.
However, due to the interaction between fields 1 and 2, the energy
difference between $(B,D)$ is higher than the energy difference
between $(A,C)$, so that 
\begin{equation}
\epsilon_{2,eff,final} = E_B - E_D = \epsilon_2 - \gamma_{2,3} a_2^3 a_3  
+ \gamma_{1,2} a_1^3 a_2 
\end{equation}
and the tunneling rate of the second field via $B \rightarrow D$ is
fast.  Hence the path taken by the fields is $A \rightarrow B
\rightarrow D$. Here we have illustrated the mechanism by which the
tunneling of one field catalyzes the tunneling of the next in the
chain via the coupling between them.  We reiterate that the exact form
of the coupling is irrelevant as long as the general picture of
tunneling from bowl to bowl is successful.

{\it Constraint:} In order for the interaction term in
Eq.(\ref{eq:epsfin}) to play a role in changing the energy difference
and hence the tunneling rate, its value must be large enough relative
to the original value of the energy difference.  The tunneling rate
must be slow without the interaction term, and large when it is
important. For the specific form of the interaction term studied, we
need
\begin{equation}
\label{eq:influence}
{\gamma_{i,i-1} a_{i-1}^3 a_i \over \epsilon_i} \geq \eta ,
\end{equation}
where $\eta$ is a number whose exact value depends on the parameters
of the potential. Earlier we showed that, because of the exponential
dependence of the tunneling rate on these parameters, in particular on
the ratio $(\epsilon_i/a_i^4)^3$, for reasonable potentials the
required change in the value of $\epsilon$ is extremely small.  Hence,
we can take $\eta \sim 1/10$ as an estimate.

Throughout the paper, we will assume that the interaction couplings
are not fine-tuned, so that all $\gamma_{i,i+1} \sim O(1)$.

{\it Reheating:} 

\begin{figure}
\centerline{\includegraphics[width=5.5in]{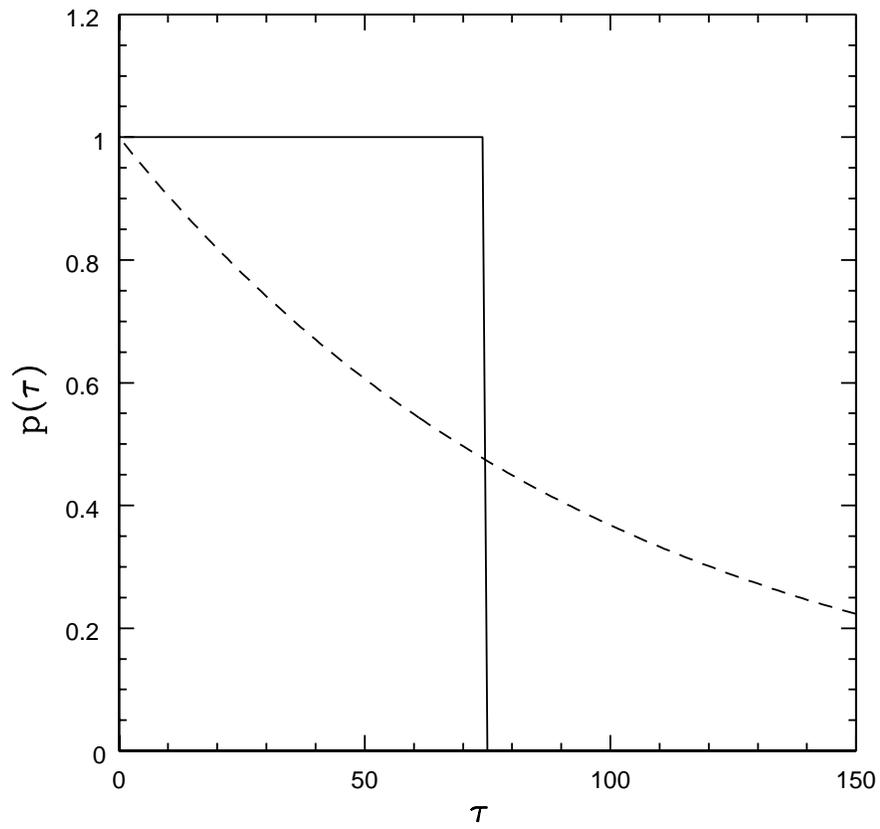}}
\caption{The probability of any point remaining in the false vacuum
  for old inflation (dotted line) and Chain Inflation (solid line).
  While the transition from slow to rapid nucleation (large to small
  $p(t)$ is too gradual for old inflation, it is virtually a step
  function for Chain Inflation, which can thus easily percolate and
  reheat.}
\end{figure}

In Fig. 2, we have plotted the probability $p(t) \propto {\rm exp}(-
{4 \pi \over 3} \beta Ht)$ of any point remaining in the false vacuum
for any of the tunneling scalar fields (other than the first one)
involved in Chain Inflation (see Eq.(\ref{eq:probds})).  Successful
inflation requires small $\beta$ initially followed by large $\beta$
at a later time, or, equivalently, large $p(t)$ initially followed by
small $p(t)$ at a later time.  In the ideal case, the transition from
large to small $p(t)$ (small to large $\beta$) would be a step
function so that percolation and reheating can easily take place.  
We have plotted the function $p(t)$ for both old inflation and chain
inflation.  While the transition from slow to rapid nucleation (large to small
$p(t)$ is too gradual for old inflation, it is virtually a step
function for Chain Inflation, which can thus easily percolate.
  
We note that, in the case of tunneling fields in Chain Inflation, this
time-dependence can be closer to the ideal of a step function even
than in the case of Double Field inflation.  Here, with a tunneler as
catalyst, there is a two state system.  When the catalyst is in its
false vacuum, there is no interaction term whatsoever and no
tunneling; with the catalyst in its true vacuum, the interaction term
suddenly turns on and tunneling is immediate.  When a rolling field is
the catalyst, on the other hand, the interaction increases more
gradually with time so that the time-dependence of $\beta$ is more
gradual.  Hence with a tunneling field as catalyst, reheating is easy
to achieve.

During each step of the chain, the tunneling rate is rapid enough that
$\beta>\beta_{crit}$ and percolation and thermalization take place.
However, the particles that are produced by the fields that tunnel
during the early stages of inflation are inflated away by the
subsequent e-folds of inflation.  Only the last field (or two) that
tunnel are relevant for reheating.  In order to reheat to a high
enough temperature to allow ordinary nucleosynthesis to take place, we
require a reheating temperature of at least 10 MeV (this is the
absolute minimum temperature one could possibly imagine).  Another
lower bound on the reheating temperature arises from baryogenesis; as
the mechanism for baryogenesis is currently not yet understood, we
allow flexibility on this value.  Hence the energy difference between
vacua of the last tunneling field $\epsilon^{1/4} \geq 10$MeV.  On the
other hand, we also require
\begin{equation}
V_{tot} < m_{pl}^4
\end{equation}
so as to stay within the bounds of applicability of ordinary effective
quantum field theory.  Thus, e.g.,for 1000 fields, we must take
$\epsilon^{1/4} < m_{GUT} \sim 10^{16}$GeV.  Hence, for all the fields in
the chain, we take
\begin{equation}
{\rm 10} {\rm MeV} \leq \epsilon^{1/4} \leq 10^{16} {\rm GeV} .
\end{equation}

\subsection{Simplest Model: Single Chain 
in which all fields have the same parameters:}
 
In the simplest version of the chain, we take the parameters of the
potential to be the same for every field other than the first one in
the chain, so that $\epsilon_i \equiv \epsilon$ and $a_i \equiv a$ for
all $i>1$.  The interaction energy between any two of the fields is
$V_{int}$.  We also take $\epsilon \sim a^4$, the most natural choice
for these parameters.

\begin{figure}
\centerline{\includegraphics[width=5.5in]{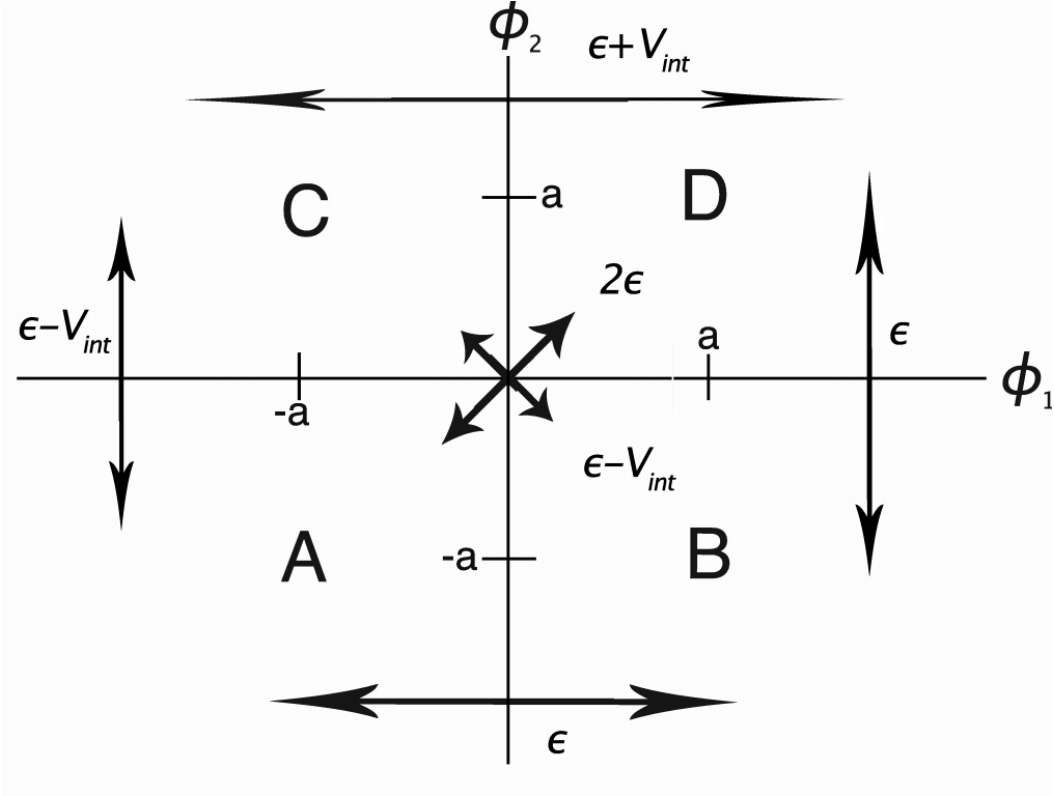}}
\caption{The four bowls (minima) in the multidimensional potential for two
  fields $\phi_1$ and $\phi_2$ for the case of identical parameters
  for the fields.  Energy differences between bowls are labeled in
  the figure.  The length of the arrows gives a rough indication of
  the amplitude of the ratio of (the energy difference to
  distance$^4$) between any two bowls. This quantity determines
  the tunneling rate.  The system chooses the path with the longest
  arrow, since its tunneling rate is the highest.  }
\end{figure}

Figure 3 illustrates the four bowls (minima) and energy differences
between them, for two fields $\phi_1$ and $\phi_2$ for this simple
case.  Using Table 1 and simplifying to the case of equal parameters
for all fields, we can evaluate the energies of the four bowls and
their energy differences The energies of the bowls are
\begin{eqnarray}
& E_A = 2 \epsilon + V_{int} \\
& E_B = \epsilon + V_{int} \\
& E_C = \epsilon + 2 V_{int} \\
& E_D = V_{int} 
\end{eqnarray}
and the energy differences between bowls are
\begin{eqnarray}
& \Delta E_{AB} = \epsilon \\
& \Delta E_{BC} = \epsilon - V_{int} \\
& \Delta E_{CD} = \epsilon + V_{int} \\
& \Delta E_{BD} = \epsilon \\
& \Delta E_{AD} = 2 \epsilon \\
& \Delta E_{AC} = \epsilon - V_{int} .
\end{eqnarray}

The highest energy bowl is A, where both fields are in their false
vacua.  The next highest energy bowl is C, where the second field is
at its true vacuum but the first field is still in its false vacuum.
The third highest energy bowl is B, where the first field has tunneled
to its true vacuum but the the second field is still stuck in its
false vacuum. The lowest energy bowl is D, with both fields in their
true minima.  No matter what path the system chooses, the total energy
difference between A and D is the same.

The system chooses the path 
\begin{equation}
A \rightarrow B \rightarrow D ,
\end{equation}
so that field 1 tunnels first and in so doing catalyzes field 2 to
tunnel.  Here the path $A \rightarrow C$ is suppressed while the path
$A \rightarrow B$ is fast. This is achieved because the tunneling rate
is slow with energy barrier $\epsilon - V_{int}$ (appropriate to $A
\rightarrow C$) but fast with energy barrier $\epsilon$ (appropriate
to $A \rightarrow B$).  In this simple case of equal parameters for
all potentials, we have
\begin{equation}
\epsilon_{eff, initial} = \epsilon - V_{int} \,\,\,\,\,\, 
{\rm (slow \, tunneling)} 
\end{equation}
and 
\begin{equation}
\epsilon_{eff, final} = \epsilon \,\,\,\,\,\, {\rm (fast \, tunneling)} .
\end{equation}
Since $\epsilon \sim a^4$ is near the border from
slow to rapid tunneling, the model works successfully if we choose
parameters in this range.

During each stage of inflation, one of the fields tunnels to its true
vacuum, so that the total energy of the system drops by an amount
$\epsilon$.  For successful reheating, the number of e-folds of
inflation $N_n$ at each stage 
cannot exceed (see Eq.(\ref{eq:ncrit}))
\begin{equation}
N_n < N_{crit} = 1/3 ,
\end{equation}
or, equivalently, $\beta_n > \beta_{crit}$, 
where subscript $n$ refers to the $n$th stage and $n \geq 0$.
The universe expands by a fraction of an e-folding
during each stage.  If we add up all the e-folds from all the stages, 
the total amount of inflation must satisfy
\begin{equation}
N_{tot} = \sum_n N_n \geq 60 .
\end{equation}
Hence, there must be at least 180 stages of inflation; i.e., the
number of fields $q$ must be at least 180.

During any one of the stages, the number of e-folds 
of inflation is given following Eq.(\ref{eq:ni}) as
\begin{equation}
\label{eq:numef}
N_n = \int H_n dt \sim H_n \tau_n = {3 \over 4 \pi}
{H_n^4 \over \Gamma_n} ,
\end{equation}
where
\begin{equation}
\label{eq:hn2}
H_n^2 = {8 \pi \over 3 m_{pl}^2} q_n \epsilon
\end{equation}
and following Eq.(\ref{eq:lifetime}),
the lifetime
$\tau = {3\over 4\pi} {H_n^3 \over \Gamma_n}$.

Here, $q_n$ is the number of fields still in their false vacua at
stage $n$.  Since some tunnel at every stage, $q_n$ drops as time goes on,
as does the number of e-folds per stage $N_n$.
At first all $q$ fields are in their
false vacua, so that $q_n = q$ and the total energy is $q \epsilon$.
During each stage, the total energy drops by an amount $\epsilon$,
and $q_n \rightarrow q_n - 1$.
Using Eqs.(\ref{eq:tunrate}), (\ref{eq:eucact}), and
(\ref{eq:hn2}) in Eq.(\ref{eq:numef}),
in the thin wall limit, the number of e-folds at a given stage is then
\begin{equation}
\label{eq:nstage}
N_n = {3 \over 4 \pi} \left({8 \pi \over 3}\right)^2
q_n^2 {\epsilon \over m_{pl}^4} 
{\rm exp}(+64{\pi^2 \over 6} \lambda^2 {a^{12} \over \epsilon^3}) .
\end{equation}

Thus we find that the amount of inflation $N_n$  with $q_n$  multiple fields 
scales as $q_n^2$ times the amount of inflation $N_*$ with a single field
in the absence of any other fields, 
\begin{equation}
\label{eq:q2}
N_n = q_n^2 N_* .
\end{equation}

Other than the first field to tunnel, the fields remain in their false
vacua during several stages.  
Although the length of each stage is a fraction of an e-foldlng,
the field that tunnels has already been in its false vacuum
for a much longer time.  For example, the third field to tunnel
has remained in its false vacuum during the first three stages.
In fact some of the later fields remain in their metastable minima 
for 60 e-folds.  Thus we choose
parameters for the potentials that would allow the metastable minima
to be stable for a very long time, more than 60 e-folds.  In each stage, 
the potential of one of the fields $\phi_n$ is modified so as to
cause it to rapidly tunnel; as we've seen $\epsilon_n$ only needs to
change by a very small amount in order for tunneling to happen.  The
length $N_n$ of that stage is set by the modified height and width of the field $\phi_n$ 
whose values are changed to their new values by the previous element in the
chain.

\subsubsection{The First Field}

The first member of the chain must make the transition from false to
true vacuum spontaneously, as it is not catalyzed to tunnel by any other
fields.  There are three possibilities for this first field.

(1) One possibility is that the first field to tunnel has (slightly)
different parameters than the rest, so that it tunnels rapidly on its
own, without assistance from coupling to the other fields.  All the
subsequent fields, on the other hand, can have exactly the same
parameters, and we choose them to have long-lived metastable minima
until the interactions change the value of $\epsilon_{eff}$.  The
parameters of the first field do not need to be very different from
the rest, due to the extreme (exponential) sensitivity of the
tunneling rate to the parameter values.

We note that this ``first'' rapid tunneler doesn't have to be the
first member in the chain. If there are 1000 links in the chain, this
rapid tunneler can be number 500 in the chain. Once this field
spontaneously tunnels, there is still sufficient inflation (60
e-folds) from fields 501-1000 in the chain.

(2) The first field in the chain can be a rolling field which produces
but a fraction of an e-fold of inflation.  Since it need not be
responsible for much inflation, nor for density fluctuations, its
potential can be arbitrarily steep.  Its parameters need not be
fine-tuned, unlike the case of slowly rolling fields in most rolling
models of inflation.  In fact, this first roller can be followed by
many more than 60 e-folds of inflation so that any information about
its properties is inflated way outside our horizon.

(3) We can imagine that the first field is a tunneling field with a
very slow nucleation rate. In fact, it can have exactly the same
parameters as all the fields.  Then the first field sits in its
metastable vacuum for a very long time before tunneling.  In other
words, it behaves much like the single scalar field in the original
old inflation model.  Hence this first field has $\beta << 1$ at all
times, and does {\it not} percolate and thermalize.  Instead there are
isolated Swiss Cheese bubbles of this first field, i.e., empty bubbles
of true vacuum.  It is only inside these empty bubble interiors (not
in the predominant sea of false vacuum) that the chain of phase
transitions continues.  The interactions between the first field and
the subsequent fields only become nonzero inside the bubble interiors
of this first field, so that catalysis of further phase transitions in
the chain only takes place inside the bubbles of true vacuum.  Hence,
our observable universe must live inside a large bubble produced by
this first trigger field.  Reheating of this large bubble is not a
problem, as it can easily happen due to the tunneling and percolation
of subsequent fields in the chain.

There is another issue one must consider if the universe lives
inside a single true vacuum bubble of the first tunneling field.  
The interior of the first bubble
is an infinite open universe.  The appropriate time slices are
hyperbolic, determined by slices of constant value of the field.  The
interior has negative curvature, so that the Friedmann equation
becomes
\begin{equation}
\label{eq:negcurv}
H^2 - {1 \over a}^2 = {8 \pi \over 3 m_{pl}^2} V_{tot} .
\end{equation}
Another sixty e-folds of inflation, subsequent to the production of
this trigger field bubble, are required in order to inflate away this
negative curvature.  This requirement can easily be satisfied by the
many subsequent tunneling fields in the chain, each of which
contributes to the total amount of inflation.

In sum, the first field in the chain can either be a spontaneous
tunneler, a rolling field, or a slow tunneler.  The phase transition
of the first field is followed by many e-folds of inflation which
erase any unwanted signatures from the first field.

\subsubsection{Example}

As an example, let us consider $q=10^4$ fields.  We will
allow the universe to inflate by 0.1 e-folds per stage (for the first
thousand or so stages).  Thus the universe inflates 0.1 e-folds before
the first field tunnels, then another 0.1 e-folds from that time until
the second field reaches its true minimum, then another 0.1 e-folds
until the third field reaches its true minimum, etc., with enough
stages to obtain a total of at least 60 e-folds.

In order to obtain 0.1 e-folds in the first stage, the number of
e-folds that would be obtained by the first tunneling field (if the
other fields were not present) is $N_* \sim 0.1/q^2 \sim 10^{-5}$.
After each stage, the number of fields participating in the inflation
decreases by 1.  In the $n$th stage, the number of e-foldings is thus
\begin{equation}
N_{n} = (q-(n-1))^2 10^{-5} .
\end{equation}
Thus, for the first 1000 stages, $N_n \sim 0.1$.  By this time there
are already a total of 100 e-foldings (adding up over the first 1000
stages).  Eventually enough fields have tunneled to their minima that
the number of e-folds per stage become very small. At the 10,000
stage, only $10^{-5}$ e-folds result. However, the total number of
e-folds, summed over all the stages, easily exceeds 60.

We can now ask how reasonable the parameters would be to obtain the
above scenario.  To honestly do so would require solving for tunneling
in the thick wall limit, as will
be considered in a later paper.   Typically tunneling is suppressed
in the thin wall limit, but we will discuss here a case where it is not
for illustrative purposes.
To tunnel in a thin wall limit requires two things.  First, 
the field should tunnel rather than slow roll; i.e.  
the energy difference between vacua should be less than the barrier height
$\epsilon<V_0$.  For the asymmetric double well this requirement becomes
\begin{equation}
\label{eq:TC}
{\lambda \over 2}{a}^4>\epsilon.
\end{equation}
Second, the thin wall condition must be satisfied.  Following Coleman 
\cite{coleman2}, the thin wall condition for an asymmetric double well is
\begin{equation}
\label{eq:TWC}
2\lambda a^4>>\epsilon.
\end{equation}
As an example, to satisfy the above conditions, we will take
${\lambda}a^4 / \epsilon=5.$

The number of e-folds due to a single tunneling event in the thin
wall limit is determined by Eq.(\ref{eq:nstage}).
It is indeed possible to find values for which
one can both tunnel quickly and satisfy the thin wall conditions.  
We note that a larger value of $\lambda$ allows for a quicker transition
and also fixes the minimal value of $a/\epsilon^{1/4}$ to satisfy
the above conditions. 

 We will now consider an example to illustrate that even in this 
unrealistic case,  fairly reasonable parameters
 can be used.  For $N_n = 0.1$ with $q=1000$ and for inflation at
the GUT scale $\epsilon = 10^{16}$GeV, we find $a/\epsilon^{1/4} =
0.26$ is required (for the parameters including the effects of coupling
to a prior field).    Similarly, for inflation at the TeV scale
$\epsilon = $TeV, we find $a/\epsilon^{1/4} = 0.47$ is required.  
In addition, to go from a slow tunneling regime (1000 e-foldings) 
to a fast regime (0.1 e-foldings) requires a 2\%  or a 3\% change of 
$a/\epsilon^{1/4}$ for respectively TEV and GUT scale fields. 
Despite the artificial nature of the thin wall limit, 
these values are quite sensible.   

\subsubsection{Caveat}

Typically, tunneling is suppressed in the thin wall limit
for any theory (we thank Erick Weinberg for pointing this out to us),
though we have discussed above some parameter ranges in which thin
wall tunneling does take place.  More generally,
a more accurate calculation of tunneling rates must be performed
numerically, in the the thick wall limit, where the energy difference
$\epsilon$ is no longer much smaller than the barrier height,
following the work of Adams \cite{fca}.  Of course, in the case
where $\epsilon$ is greater than the barrier height, there is no tunneling
at all and the field simply rolls down the potential.
Then too little inflation would ensue.  If one
were to allow a distribution of parameters, even one peaked about
$\epsilon = a^4$, then it is plausible that a significant fraction of
the fields would remain in their false vacua long enough for the model
to work.  In particular, as we will see in Section 4.2, if there are
different paths taken by neighboring patches of the universe, those
regions that inflate more end up much larger than those that inflate
little, and our observable universe is more likely to end up within
the larger patch.  It is only the extreme case where the rapidly
tunneling fields couple to all the intrinsicallly slowly tunneling
fields and cause them to tunnel rapidly as well (or roll), that is dangerous.
Below we will also consider interactions that act in the direction of
slowing down the tunneling, which would assist in this case.  [These
  interactions could exist initially and disappear in time.]  However,
in general, we do want to warn that there is a range of parameter
space for which the potential is not fine-tuned, and yet Chain
Inflation may not work.  For example, if the coupling parameter
$\gamma_{i,i-1}$ between the fields were large, then the resulting tunneling
could become far too rapid (or, depending on the details of the coupling could
become far too slow); note that we have assumed all couplings are 
naturally $O(1)$.  There is one basic requirement on a
multidimensional potential in order to obtain successful Chain
Inflation: there must be enough long-lived fields, in the sense that
they remain in their false vacua for a significant fraction of an
e-folding before being catalyzed to tunnel to their vacua.  Then
sufficient inflation will result.

\subsubsection{Issues: Would the Path skip a Link in the Chain?}

One might ask whether or not the path taken by the
system on the way to its ground state would preferentially skip
elements in the chain.  In a landscape picture, the path could jump
directly to a bowl that is lower in energy by several $\epsilon$; this
alternate route would be dangerous if it were the quickest path
towards the ground state, as fewer e-folds of inflation would result.
We can show that the system does not skip steps for the case
where all the potentials have identical parameters; the more
general case will be considered in the next section.

Let's consider two fields $\psi_1$ and $\psi_2$, each with
a double-well potential of the same parameters (same $\epsilon$ and
$a$).  Following the discussion above,
in the two-dimensional field space, there are four bowls:
(1) at point A, $\psi_1 = \psi_2 = -a$, (2) at point B,
$\psi_1 = +a$ and $\psi_2 = -a$, (3) at point C, $\psi_1 = -a$
and $\psi_2 = +a$, and (4) at point D, $\psi_1 = \psi_2 = +a$.
The system starts out at point A, where both fields are in their false vacua.
Then, one can ask the question of whether
it is faster to proceed from $A \rightarrow B \rightarrow D$ or
directly from $A \rightarrow D$.  The latter path would be dangerous
if it were the fastest, as a path that proceeds quickly to the ground
state might not inflate enough.  As a reminder, the tunneling rate
scales as $\Gamma \propto {\rm exp}[-F (a^4/\epsilon)^3]$ where
$F=\pi^2 \lambda^2/6$.  In comparing the path ($A \rightarrow B$) vs.
($A \rightarrow D$), we see that we need to compare $\epsilon$ vs.
$2 \epsilon$ and $a$ vs. $\sqrt{a^2 + a^2} = \sqrt{2} a$, so that
the relevant comparison is between 
$a^4/\epsilon$ vs. $2 a^4/\epsilon$.  We can now compare the
time it takes to follow the two paths:
\begin{equation}
{\rm Time} (A \rightarrow D) \propto
{\rm exp}[F 8 (a^4/\epsilon)^3] >> 2 \times {\rm exp}[F (a^4/\epsilon)^3]
\propto {\rm Time} (A \rightarrow B) + (B \rightarrow D) .
\end{equation}
Hence the preferred path is to tunnel in the direction in field space
in which the energy decreases by a single unit of $\epsilon$ at a
time; i.e., which can be thought of as the tunneling of a single
field.

\subsection{Generalizing the Model}

Chain Inflation is a general phenomenon in which the system
of scalar fields in a multidimensional potential tunnel from bowl
to bowl.  Our particular choice of couplings allow simple calculations
so as to obtain reasonable estimates. However, the coupling
certainly need not be of the form presented here.  Indeed, all
the fields could couple to each in other in a complicated fashion 
as long as the couplings preserve the bowl-type of structure in a
multidimensional landscape.

Previously we considered the simplest single chain model in which all
fields have the same parameters.  In this section we  discuss some
simple  generalizations of this
model which would also work.  We restrict our discussion to double-well
potentials, but the results should generalize to other choices
of potentials.

\subsubsection{Arbitrary parameters, a number of chains}
First, let us consider the effects of keeping the same form of the
potential and the interactions, but allowing a variety of parameters
$a_i$ and $\epsilon_i$ for the different fields.  In addition, we will
allow for the existence of a number of chains. The first field could
couple to a number of secondary fields (next in line in a chain after
the first field).  Then each of the secondary fields couples to a
number of tertiary fields, etc.  Some subset of the secondary fields
could also couple to each other.

In one patch of real space, the fields follow a single path in
(multidimensional) field space; i.e., at each instant, there is a
single vacuum expectation value for each of the fields $\langle \phi_1
\rangle, \langle \phi_2 \rangle ...$.  Of the many choices of
direction in field space, e.g. the variety of possible secondary
tunneling fields, the system only chooses one.  At each stage, the
path en route to the ground state that would be chosen would be the
fastest one.  For example, if the first field couples to ten others,
then the system chooses the direction in field space with the fastest
tunneling rate to a lower energy. The system ``chooses'' the field
that tunnels the most quickly.  The other nine fields to which the
first field couples become irrelevant.

Here we address three issues that could arise for different choices of
parameters in the potentials: (i) fields which tunnel too slowly, (ii)
fields which tunnel too rapidly, and (iii) interactions which are so
large that the chain sequence is modified.

{\it Tunneling too slowly:} Due to the variety of parameters, there
might be some fields which, even after their interactions turn on,
still are stuck in their false vacua.  In the language of the
landscape, even at field values in the multidimensional field space
where the interaction of the given field with the preceding member of
the chain is nonzero, still the tunneling rate of the field is
extremely slow.  In that case, the path that the system takes en route
to the ground state will simply avoid that direction.  As long as one
of the secondary fields is able to tunnel to a lower minimum, the
expectation values of the fields will choose the path that has a seris
of rapid tunnelers.  In a landscape picture, the system will take
the path of least resistance to the ground state.  In addition,
generically there are multiple couplings between fields, and it is
likely that one of these couplings will suffice to induce the
(otherwise slowly tunneling) field to quickly tunnel.

Even if the path that is chosen does include a very slow tunneler
in the middle of the chain, as long as there are sixty e-folds 
subsequent to that slow tunneler, any resultant negative curvature
will get washed out by the subsequent inflation.

{\it Tunneling too rapidly:} A new problem could arise, however.
Somewhere down in the chain, e.g.  the fiftieth field in the chain,
could have parameters (e.g. a large enough $\epsilon$) such that it
tunnels spontaneously, without being catalyzed by any other field.  If
we think of the single chain as a series of dominos, then this case
corresponds to the 50th domino falling over on its own.  If there are
100 dominos in total, then the first set of 50 and the second set of
50 will fall over at the same time. It will take half as long for all
of them to fall over, i.e., only half as many e-foldings of inflation
will result, so the overall vacuum
energy will drop down by an amount $\epsilon$ twice as quickly.
However, sufficient inflation should still result as long as (i) there
are significantly more than 60 steps in the chains and (ii) relatively
natural values of parameters for the potentials are chosen, i.e.,
$\epsilon \leq a^4$ for most of the fields, so that most fields do not
tunnel spontaneously (see the discussion in Section 4.1.3).

At different spatial points in the universe, it is possible that the
system chooses different paths in field space.  For example, of the
ten possible secondary paths in the above example, it is possible that
two of them have the same tunneling rate and are equally likely to be
chosen.  Then, in the end, two different patches of the universe that
followed slightly different paths could end up at different ground
states, both with $V=0$ but with different field values.  Topological
objects such as domain walls could be formed in between these
different patches.  Of all the possible paths to the ground state,
some may be too rapid, in the sense that the total number of e-folds
is too small (e.g. because of the effect discussed in the last
paragraph).  Then clearly we do not live in a patch of the universe
that took such a steep, rapid path.  Since this under-inflated patch
of the universe is much smaller than the patches that did inflate, it
is quite reasonable that our observable universe lies within the much
larger regions that did inflate sufficiently.

{\it Too large interactions:} Our treatment in Eq.(\ref{eq:vtot}) has
assumed that one can break up the potential into a set of asymmetric
double well potentials with interactions. However, if the interactions
are very large, $V_{int} > V_i(\phi_i)$ for one of the fields, this
basic picture is not quite right. If two subsequent fields in the
chain have very different energy scales, then the order in which
fields tunnel in the chain may change.  For example, if a field with a
TeV scale potential is followed in the chain by a field with a GUT
scale potential, the GUT scale potential will override the TeV one and
will tunnel first.

To illustrate this effect, let us consider two of the fields in the
system, 1 and 2, as studied in Table I and the accompanying
discussion.  Now let us take the first field to be characterized by a
TeV scale, $\epsilon_1 \sim a_1^4 = TeV^4,$ while the second field is
characterized by a grand unified (GUT) scale $\epsilon_2 \sim a_2^4
\sim (10^{16} {\rm GeV})^4. $ We will also assume $a_0, a_3 <<
m_{GUT}$.  The four bowls (minima) now have energies
\begin{eqnarray}
& E_A = m_{GUT}^4 + TeV^4 +a_0^3 TeV , \\ 
& E_B = m_{GUT}^4 + TeV^3 m_{GUT} , \\
& E_C = TeV^4 + a_0^3 TeV + m_{GUT}^3 a_3 , \\ 
& E_D = m_{GUT}^3 a_3 .
\end{eqnarray}
The system starts out in bowl A, where both fields are in their false
vacua.  Ordinarily field $\phi_1$ would tunnel first and the system
would then move to bowl B.  Here, however, the interaction is so big
that the energy difference between bowls A and B is negative,
\begin{equation}
\Delta E_{AB} = E_A - E_B  = TeV^4 - TeV^3 m_{GUT} + a_0^3 TeV < 0 ,
\end{equation} 
i.e.
bowl B is at a {\it higher} energy than bowl A. Clearly the fields
will not tunnel from A to a state B of higher energy.
Instead, the
second (larger) field $\phi_2$ tunnels first.
The energy difference between bowls A and C is 
\begin{eqnarray}
&E_{AC} = m_{GUT}^4 - m_{GUT}^3 a_3 
& \sim m_{GUT}^4 \,\,\,\, ({\rm for} \,\,\, a_3 \ll m_{GUT}) .
\end{eqnarray}
Thus the system preferentially takes the path $A \rightarrow C
\rightarrow D$. The resultant number of e-folds is not significantly
different from the number obtained in the normal sequence of the chain
as discussed previously.  If there is a string of mismatched scales,
one may lose some e-foldings but with the large number of fields there
should still be no problem getting enough inflation.

{\it Getting the right amount of inflation:} The chain model works as
long as potentials are not fine-tuned.  For the double-well example,
as long as $\epsilon \sim a^4$ for a large fraction of the fields,
they are on the border between remaining in metastable vacua and rapid
tunneling; in addition, they are able to influence one another to
tunnel.  If all the fields have $\epsilon >> a^4$, then the system may
reach the ground state without inflating enough.  If, on the other
hand, all the fields have $\epsilon << a^4$, then one might worry that
they all remain in their false vacua too long, although their
couplings to other fields in that case will probably induce them to
tunnel and percolate.  [If there is some discrete symmetry that forces
  $\epsilon=0$, so that both minima in the double-well potential are
  degenerate, obviously there is no tunneling at all.]

\subsubsection{More general interactions}

We have considered a very simple form of the interaction, $V_{int}
\propto (\phi -a)$, where $\pm a$ are the field values at the minima.
More generally, the interaction could be of the form
\begin{equation}
V_{int} \propto (\phi \pm b)^p 
\end{equation}
where $b$ is an arbitrary field value and $p$ is an arbitrary power.
The fact that the value of $b$ does not necessarily equal the value of
$a$ does not qualitatively change the model.  The amount by which
$\epsilon_{eff}$ differs from $\epsilon$ due to the interaction has to
be recalculated, but the effect of causing a transition from slow to
rapid tunneling still takes place (one must of course check that the
interaction is strong enough to change $\epsilon$ by 1\%.

Due to the possibility of the opposite sign in front of $b$, the
interaction could have the opposite effect from what we want. It could
inhibit the field from tunneling, rather than catalyze the tunneling.
One could imagine that half of the interactions felt by a given field
would serve to increase the tunneling rate, and half to decrease it.
As different trigger fields tunneled to their minima, they could act
in either direction. However, the symmetry is broken by the fact that
the field only has to be induced to tunnel once.  Once it has
tunneled, then the sign of future interactions is irrelevant.  Also,
the system will take the path of least resistance, i.e., the path
along which the interaction serves to speed up rather than slow down
the tunneling.

One could also imagine an interaction which is nonzero only when
catalyzing field is still in its false vacuum; for example, the
catalyzing field could prevent tunneling of another field early on
(due to the interaction) and then play no further role once it reaches
its minimum.

\subsubsection{Thermal Activation}

One could imagine that, depending on the values of the parameters, as
one of the fields in a stage of Chain Inflation successfully reheats,
it thermally activates one of the other fields to go over the top of
its barrier. Heretofore we have not considered the effects of such
thermal activation, which might be interesting to pursue.

\subsubsection{Mixing Rollers and Tunnelers}

In the paper we have restricted the discussion to a series of
tunneling fields. However, there is no reason to do so.  In the
landscape, one could imagine a mixed succession of rolling fields and
tunneling fields.  The path chosen towards the ground state could
involve rolling for a while, followed by a tunneling event, followed
again by a period of rolling down a potential.  Such a mixed chain
would be interesting to study further.

\section{Variants}

Here we briefly comment on three variants of tunneling inflation with
multiple fields which are alternatives to the chain we have been
discussing: (i) a large number of uncoupled tunneling fields which
tunnel simultaneously, (ii) two-tunnelers, and (iii) the case where
the first tunneling field in the chain inflates sixty e-folds and then
catalyzes a large number of fields to simultaneously tunnel and
percolate.  We will see that all these ideas are fatally flawed if one
restricts the discussion to multiple tunneling fields only.

However, we emphasize that the third idea may succeed if the trigger
field is a rolling field (as in the case of the Double Field model)
which, once it reaches its minimum, catalyzes the simultaneous
tunneling of a large number of secondary fields.  This idea will be
discussed in future work.

\subsection{Multiple uncoupled tunneling fields}
Here we consider a large number $q$ of uncoupled tunneling fields,
analagous to the large number of uncoupled rolling fields in assisted
inflation\cite{liddle}.  The number of e-folds obtained from any one
of the fields is given using Eq.(\ref{eq:ni}) as,
\begin{equation}
N_i \sim {8 \pi \over 3 m_{pl}^2} \sqrt{\epsilon_i} \tau.
\end{equation} 
If each field produces a small amount of inflation $N_i<<1$, the total
number of e-folds from all the fields together is $N \sim q^2 N_i$
(see Eq.(\ref{eq:q2})).  One can easily imagine obtaining sufficient
inflation, $N > 60$.  However, then the parameter $\beta \sim {1 \over
  q^2} \beta_i << 1$ is a constant value independent of time, and
never reaches $\beta > \beta_{crit}$ required for perocolation and
thermalization.  This model is equivalent to inflating with a single
field of energy density $\rho = q \epsilon$ with constant nucleation
rate.  Hence the required criteria of sufficient inflation (small
$\beta$) followed by reheating (large $\beta$) can never be achieved
for constant $\beta$.  In short this model fails for the same reason
old inflation does.  Thus, although the model of assisted inflation,
with multiple uncoupled rolling fields, can succeed, the equivalent
model with multiple uncoupled tunneling fields fails.

\subsection{Two Tunnelers}

One might propose an alternative to the Double Field model, in which
there are two tunneling fields instead of one roller and one tunneler.
Here the first tunneling field serves as the 'trigger' field to
catalyze the tunneling of the second field.  At first, both fields are
in their false vacua.  Then, when the first field tunnels to its true
minimum, it catalyzes the second field to tunnel as well.  In the
discussion of the Double Field model above, in Eq.(\ref{eq:vint}) one
could simply replace $\psi^3$ with $(\phi_1+a)^3$, where $\phi_1$ is
the trigger field, whose potential has a false vacuum at $\phi_1=-a$
and a true vacuum at $\phi_1 = + a$.  Hence the interaction term only
turns on once the trigger field has tunneled to its true vacuum. At
that point, the energy difference between minima in
Eq.(\ref{eq:epsiloneff}) of the second field becomes so large that its
tunneling rate becomes very large, allowing bubbles of its true vacuum
to nucleate simulataneously throughout.

In this two tunneler model, the trigger field must remain in its false
vacuum for at least sixty e-folds, because once it tunnels to its true
vacuum, inflation is over.  Hence, this trigger field must have $\beta
<< 1$.  It does {\it not} percolate and thermalize.  Instead, there
are isolated Swiss Cheese bubbles of this first field, i.e., empty
bubbles of true vacuum of the trigger field.  Since the interaction
term only becomes nonzero for the bubble interiors from the trigger
field, the secondary field only undergoes the phase transition inside
these trigger field bubbles.  All the bubbles from secondary field
must live inside a large bubble produced by this first trigger field.
The idea would be to reheat the interior of one of the bubbles of the
trigger fields with the energy density from the bubble collisions of
the secondary tunneler.  Since the secondary tunneler does have a
time-dependent nucleation rate, due to interaction with the trigger
field, the bubbles from the second tunneler can easily percolate and
thermalize. If $\epsilon$ is large enough for this second tunneler,
there is no problem reheating the inside of the big bubble from the
first tunneler.

However, in the case where the trigger field is a tunneling field, the
universe we live in today cannot have originated inside the interior
of a true vacuum bubble of the first tunneler.  The interior of the
bubble is an infinite open universe with negative curvature as in
Eq.(\ref{eq:negcurv}).  Another sixty e-folds of inflation, subsequent
to the production of this trigger field bubble, are required in order
to inflate away this negative curvature.  But in the model where the
secondary field tunnels right away once the trigger field bubble comes
into existence, these 60 e-folds do not take place. Hence the
two-tunneler model fails\footnote{In the chain picture considered
  previously, where there is a series of secondary fields, there was
  sufficient inflation following the trigger tunneling event.}.

\subsection{Single Trigger Field Catalyzing Multiple Secondary Fields
to Simultaneously Tunnel}

Here we consider inflation with multiple fields which are all coupled
together in such a way that, once the first one reaches its true
minimum, it catalyzes all the rest to tunnel immediately.  This model
is similar to the two-tunneler model discussed above, but with the
single secondary field replaced by multiple secondary fields which all
tunnel at once.  We will see that this model suffers from the same
problems as the two-tunneler model above.

In order for this model to work, the first field must remain in its
false vacuum for sixty e-folds; once it tunnels, all the other
secondary fields simultaneously and very quickly tunnel, thereby
ending the inflation.  One could imagine, e.g., that the potential for
each of the fields is a double well as in Eq.(\ref{eq:doublewell}),
with interaction terms of the form
\begin{equation}
\label{eq:simmultunn}
V_{int,i} = - {1 \over 16}\gamma_i (\phi_i - a_i) (\phi_1 + a_1)^3 .
\end{equation}
Here $\phi_1$ is the trigger field.  As in the Double-Field model of
Adams and Freese, once the trigger field reaches its minimum,
$\epsilon_i \rightarrow \epsilon_{eff,i}$ and all the fields tunnel at
once and successfully reheat.  Here, $\epsilon_{eff,i} = \epsilon_i +
{1 \over 8} 
a_i \gamma_i (\phi_1 + a_1)^3$.  Once the trigger field reaches its
minimum at $\phi_1 = a_1$, $\epsilon_{eff,i} = \epsilon_i + 
\gamma_i a_1^3 a_i$.  Due to the abrupt change in tunneling rate for
all the secondary fields from very slow to very fast, these fields can
easily nucleate bubbles of true vacuum throughout the universe and
hence percolate and thermalize.

However, as in the two-tunneler model, the interiors of bubbles from
the initial trigger field are problematic.  The trigger field must
remain in its false vacuum for at least sixty e-folds, because once it
tunnels to its true vacuum all the other fields tunnel as well, and
inflation is over.  Hence, this trigger field must have $\beta << 1$,
it does not percolate, and empty bubbles result from this first
tunneling.  The interaction term in Eq.(\ref{eq:simmultunn}) is only
nonzero inside these empty bubbles, so that the secondary fields only
tunnel to their true vacua inside the trigger bubbles.  Again,
although the interior of the big bubble can successfully reheat due to
the percolation and thermalization of the secondary bubbles, the
interior does not look like our universe.  It is an infinite open
universe with negative curvature.  Another sixty e-folds of inflation,
subsequent to the production of this trigger field bubble, are
required in order to inflate away this negative curvature.  But in the
model where all the secondary fields tunnel right away once the
trigger field bubble comes into existence, these 60 e-folds do not
take place. Hence, one must return to the chain picture that we have
discussed as the primary model of this paper, in which the first field
triggers a chain of tunnelers that approach their minima serially and
in the process give rise to another 60 e-folds of inflation.

We note a variant of this idea that avoids the problems discussed
above. The trigger field may be a rolling field. The key difference is that
the first field is rolling, as in the case of the Double Field model.
Once this rolling field reaches its minimum, it catalyzes the
simultaneous tunneling of a large number of secondary fields. If one
uses a rolling field as the initial trigger field, then there is no
issue of negative curvature.  However, one must address the issue of
parameters of the rolling field to demonstrate that one can avoid
fine-tuning.  This subject is the topic of an upcoming paper.

\section{Issues}

\subsection{Can de Sitter violations of energy conservation
send a tunneling field over the top of the barrier?}

Here we address the question of whether or not de Sitter violations of
energy conservation could render the barrier between false and true
vacuum in the double wells irrelevant, in that the energy violations
are so large that they send the field over the top of the barrier.  We
show that the effect is not important in our model.

Since the de Sitter metric has no timelike Killing vector globally
defined, there are violations of energy conservation of magnitude
\begin{equation}
\Delta E \sim \sqrt{\rho_{vac}/m_{pl}^2}
\end{equation}
where $\rho_{vac}$ is the vacuum energy density. For us, $\rho_{vac}$
is the largest at the beginning of inflation, $\rho_{vac} = q
\epsilon$ where $q$ is the number of fields.  If $\Delta E > {\rm
  barrier \, energy}^{1/4}$, then the field can hop over the top of
the barrier.  For this to happen, one would need $q \epsilon >
m_{pl}^2 \epsilon^{1/2}$.  If we take the mass scale of the barrier
energy to be GUT scale, this corresponds to $\epsilon^{1/2} \sim
10^{-6} m_{pl}^2$.  Then the condition for the field to be able to hop
is never satisfied for $q<10^6$.  At potentials with lower energy scales,
it is even more difficult for energy violation to cause the field
to hop over the barrier, and an even larger number of fields would
be required.  Thus we conclude that, in the
models we have been considering, the field doesn't go over the top of
the barrier due to energy non-conservation.

\subsection{Fine Tuning}

Whereas slowly rolling models of inflation require unnaturally small
parameters, the proposed model of Chain Inflation has the attractive
feature that the potential need not be fine-tuned.

In inflationary models with a single scalar field that is slowly
rolling, the potential is characterized by its height height $\Delta
V$ (the vacuum energy of the inflaton) and width $\Delta \psi$ (the
change in field value during inflation).  For inflation to work in
this context, it must satisfy certain constraints: there must be
sufficient inflation and the amplitude of the density perturbations
\cite{star,bar} cannot exceed $\delta \rho/\rho \leq 10^{-5}$.  A
fine-tuning parameter,
\begin{equation}
\chi \equiv \frac{\Delta V }{\Delta \psi^4} ,
\end{equation}
has been introduced to examine these constraints \cite{afg}.  The most
natural value of this ratio in particle physics would be $\chi \sim
O(1)$. Instead, in order to satisfy the above two constraints, it has
been shown that this parameter must satisfy \cite{afg}
\begin{equation}  
\chi \leq 10^{-8} ,
\end{equation}
a very unnatural value.  This strongly constrains models of slow roll
inflation.  Very few inflation models (such as natural inflation
\cite{nat,natural2}) can explain this small number.

The potentials in Chain Inflation do not require fine-tuning.  The
potentials we have considered have $\epsilon \sim a^{1/4}$, i.e., the
parameter $\chi \sim 1$.  This is an advantage of the Chain Inflation
model.

As discussed in Section 4.1.3, however, the multidimensional potential
in Chain Inflation must have the right parameters to allow the fields
to remain in their false vacua for a long enough time to obtain at
least sixty e-folds in all.

\subsection{Density Fluctuations}

There are several remaining issues for future work. The most important
is the question of generation of density fluctuations.  One source of
perturbations would be the fact that the lifetime of a field in its
false vacuum is given only roughly by Eq.(\ref{eq:lifetime}).  In
fact, there would be a distribution of timescales about this typical
value.  Some regions of the universe would tunnel to their true minima
on a slighter shorter timescale than their neighbors. In different
patches of the universe, the phase transition could be a little ahead
or a little behind, and this timing difference can lead to density
fluctuations.  This situation is similar to the density fluctuations
in slow-roll inflation that are produced as a result of neighboring
regions rolling to their minima at slightly different times (due to
quantum fluctuations at the beginning of rolling).  As in new
inflation, here again the fluctuations that are created sixty to fifty
e-folds before the end of inflation are on length scales that today
correspond to observable structure.  It is not clear what the spectrum
of these perturbations would be, or whether they would be adiabatic or
isocurvature.  This issue must be investigated further.

In those regions which tunneled the most quickly, the resulting
bubbles would have more time to grow than in the neighboring regions.
In \cite{tww}, the authors worried about the ``Big Bubble'' problem
that resulted in inflation with a single field; the universe could not
have tolerated the excessively big bubbles that would arise.  Here,
however, the size difference between bubbles of any one field is given
by ${\rm exp}(N_n)$, and $N_n$ is a small number (less than 1/3).
Thus the largest bubbles cannot even be twice as large as the typical
bubbles, and would not create unduly large inhomogeneities.  They
would instead potentially create the density fluctuations that we see.
It takes an extra amount of time, ${\rm exp}[N_n/2]$, for energy in
the bubble walls to convert to particles. One finds this amount of
extra energy near the bubble walls.  One can ask whether this extra
energy can be smoothed out across the bubbles.  The particles (at this
point relativistic) move across the bubbles with the speed of light.
Hence in a Minkowski background, the particles could easily thermalize
across the interior.  However, here the background spacetime is still
de Sitter after one of the fields has tunneled to its minimum while
some of the other fields have not, and the system is not yet at its
ground state.  Hence the interior space is expanding superluminally
and the particles cannot traverse the interior.  One ends up with
spherical shells of particles. Of course, as the universe continues to
inflate, the densities of these particles is diluted to the point
where they are irrelevant.  In short, percolating bubbles in the early
stages of inflation turn into spherical shells of particles (where the
bubble wall was) that cannot thermalize, because the interior is de
Sitter and information can only traverse the bubble at the speed of
light.  However, the situation is different at the very final stage of
inflation, when the last field in the chain tunnels down to $V=0$.
The bubbles of this last stage of inflation can thermalize so that
reheating is successful.  As the system has now tunneled down to
Minkowski space, the bubble interiors are not expanding
superluminally.  The particles produced near the bubble walls can
easily traverse the entire bubble in one Hubble time and reheat the
interior.  In addition, the spherical shells of bubbles produced at
earlier inflationary stages can now also spread out throughout the
universe, again in a Hubble time.

To reiterate, the key question as regards density fluctuations is the
timing of the phase transitions in different patches of the universe,
as discussed above and as should be investigated further.

Density perturbations could result from other effects as well.  For
example, perturbations to the exact $O(4)$ symmetry of the solution to
the Euclidean action might lead to deviations in the density.
Other effects leading to perturbations (in rolling models) were 
discussed in \cite{ars} and \cite{kk}.

\subsection{Beyond Thin-Wall}

A more accurate treatment of the chain double well model discussed in
this paper would require going beyond the thin wall limit.  We are
aware that the thin-wall approximation is valid only in the case where
$\epsilon/a^4 << 1$, in which case the bounce action is really small
and there is no tunneling at all.  Hence in those cases of interest
where tunneling takes place, one should numerically solve the bounce
equation.  However, the basic picture presented in the current paper would remain
unchanged, though the detailed numerical answers might be different.
Adams \cite{fca} has previously studied generic quartic potentials in the 
thick wall limit.   The chain model as presented still results
(as long as $\epsilon < a^{4}$ so that the field is tunneling
rather than rolling).

\section{Conclusion}

In conclusion, we have proposed the model of Chain Inflation, in which
a sequential chain of coupled scalar fields can drive inflation.  We
considered a toy model of a chain of tunneling fields, each of which
catalyzes the next to tunnel to its true vacuum.  Since each tunneling
stage provides only a fraction of an e-folding, percolation of the
true vacuum bubbles and hence reheating is easily achieved. Many
fields, at least several hundred, are required in order to achieve
enough inflation.  Such a large number of fields is motivated by the
many vacua in the stringy landscape, but our model can apply to a
chain of tunnelers in any multidimensional potential.  One can think
of each tunneling event as equivalent to dropping from one bowl in a
multidimensional potential to another bowl of lower energy, until the
zero energy ground state is achieved. Chain Inflation has the
attractive feature that it relies on couplings between the fields,
which are likely to exist.  We have focused on double-well potentials
as a toy model, first with identical parameters and couplings, and
then generalized to arbitrary values.  However, the idea is quite
general, and relies only on the idea of tunneling in a
multidimensional potential from one minimum to the next, regardless of
any details of the potential.  Chain Inflation works for natural
values of parameters and couplings.  It can be successful for a wide
variety of energy scales for the potential, ranging from values as
low as 10 MeV up to a GUT scale at $10^{16}$ GeV.

\section*{Acknowledgements}
 
We would like to thank Fred Adams, Alan Guth, Jose Blanco-Pillado,
Will Kinney, Gary Shiu, Dejan Stojkovic, and Erick Weinberg for many
useful discussions.  We are particularly grateful to Jim Liu for his
invaluable contribution during many helpful conversations about the
model.  We acknowledge support from the DOE via a grant at the
University of Michigan, and thank the Aspen Center for Physics as well
as the Michigan Center for Theoretical Physics for hospitality and
support while this work was completed.

\end{document}